\documentclass[aps,prd,amsmath,twocolumn,superscriptaddress,floatfix,nofootinbib]{revtex4}
\usepackage[dvips]{color}
\usepackage{epsfig}
\usepackage{latexsym}
\usepackage{graphicx}
\usepackage{amsfonts}
\usepackage{amsmath}
\usepackage{todonotes}
\usepackage{natbib}
\usepackage{wasysym}
\usepackage{bm}
\usepackage{array}
\usepackage{enumitem} 

\usepackage{color}
\usepackage{xcolor}
\usepackage{blindtext}
\usepackage{ulem}  
\usepackage{comment}
\usepackage{hyperref}
\usepackage{pdfpages}
\usepackage{orcidlink}

\newcommand{\f}{\begin{equation}}
\newcommand{\ff}{\end{equation}}
\newcommand{\fnl}{\ensuremath{f_{\rm NL}}}

\begin{document}

\title{
Curvaton in light of the ACT results}

\author{Christian T.~Byrnes\,\orcidlink{0000-0003-2583-6536}}
\email{C.Byrnes@sussex.ac.uk}
\affiliation{Department of Physics and Astronomy, University of Sussex, Brighton BN1 9QH, UK}

\author{Marina Cort\^{e}s\,\orcidlink{0000-0003-0485-3767}}
\email{mvcortes@ciencias.ulisboa.pt}
\affiliation{Institute of Astrophysics and Space Sciences, Faculty of Sciences, University of Lisbon, 1769-016 Lisbon, Portugal}
\author{Andrew R.~Liddle\,\orcidlink{0000-0002-8164-4830}}
\email{arliddle@ciencias.ulisboa.pt}
\affiliation{Institute of Astrophysics and Space Sciences, Faculty of Sciences, University of Lisbon, 1769-016 Lisbon, Portugal}

\date{\today}

\begin{abstract}
The latest results from the Atacama Cosmology Telescope Collaboration have moved the preferred perturbation spectral index $n_{\rm s}$ closer towards one. We reanalyse constraints on the simplest version of the curvaton model, using $n_{\rm s}$ and the tensor-to-scalar ratio $r$. We show that in the massless curvaton case the model gives the same locus of $n_{\rm s}$--$r$ predictions as the power-law model $V(\phi) \propto \phi^p$, but with a different and more elegant physical interpretation. The model gives an excellent account of current observational data, both in the regime where the curvaton dominates the observed perturbations and with an admixture of inflaton-originated perturbations of up to about one-third of the total.  Addition of the newest South Pole Telescope SPT-3G D1 dataset maintains this conclusion in favor of the curvaton model. Forthcoming large-scale structure surveys, such as that of the recently-launched SPHEREx probe, could discriminate between curvaton and single-field inflation models using the bispectrum.
\end{abstract}

\maketitle

\tableofcontents        

\maketitle
\flushbottom

\section{Introduction}

The analysis of cosmological constraints by the Atacama Cosmology Telescope (ACT) collaboration, following their DR6 data release, has indicated a consequential shift upwards in the preferred values of the scalar spectral index $n_{\rm s}$, from the {\it Planck} 2018 value  of $n_{\rm s} = 0.9651 \pm 0.0044$ \cite{Planck:2018vyg} to $n_{\rm s} = 0.9743 \pm 0.0034$ \cite{AtacamaCosmologyTelescope:2025blo} (in a compilation using the DR1 data from the Dark Energy Spectroscopic Instrument (DESI) \cite{DESI:2024uvr,DESI:2024lzq,DESI:2024mwx}). The new central value is just over 2-sigma from the {\it Planck} result. While within the realm of plausible statistical fluctuations, this change has substantial implications for inflationary modelling. The higher value has been reinforced in subsequent analyses, for example the revised (v2) ACT paper \cite{AtacamaCosmologyTelescope:2025blo} includes an analysis updating to DESI-DR2 \cite{DESI:2025zgx} that yields the somewhat higher again $n_{\rm s} = 0.9752 \pm 0.0030$, while
the Dark Energy Survey Y6 release finds a yet higher value 
$n_{\rm s} = 0.977 \pm 0.003$ from an all-data compilation including {\it Planck} and ACT \cite{DES:2026fyc}.



It is worth remarking that only part of the shift comes from the ACT DR6 power spectrum data itself, which combined with {\it Planck} gives $n_{\rm s} = 0.9709 \pm 0.0038$  (ACT alone gives quite a wide range for $n_{\rm s}$, but the intersection of preferred regions with {\it Planck} in the full parameter space pulls to the higher values). Only when the ACT and {\it Planck} power spectra are combined with cosmic microwave background (CMB) lensing and with baryon acoustic oscillation constraints from DESI-DR1 \cite{DESI:2024uvr,DESI:2024lzq} does the full shift emerge \cite{AtacamaCosmologyTelescope:2025blo}. 
Combined with the very strong upper limits on the tensor-to-scalar ratio $r$ coming from BICEP/Keck \cite{BICEP:2021xfz}, the viable region of the $n_{\rm s}$--$r$ plane is becoming small as well as being displaced from its position of a few years ago.

For instance, those who invested heavily in Starobinsky inflation \cite{Starobinsky:1980te} and its variants \cite{Kallosh:2013yoa} following {\it Planck} now find themselves in the uncomfortable position of only touching the 95\% confidence region for a very high pivot $e$-foldings number of $N_*=60$ \cite{AtacamaCosmologyTelescope:2025nti},\footnote{It is worth remembering that early inflation modelers chose $60$ as an estimate for when the present Hubble radius crossed outside during inflation, whereas modern observations are specified at $k_* = 0.05 \, {\rm Mpc}^{-1}$ which crossed five $e$-foldings later \cite{Liddle:2003as}.} possibly requiring non-standard reheating \cite{Drees:2025ngb,Zharov:2025zjg,Haque:2025uri,Haque:2025uis}. Another popular option, Higgs inflation \cite{Bezrukov:2007ep}, has also suffered under this change \cite{Zharov:2025zjg,Liu:2025qca}. 
Alternatively, rather than seeking to resuscitate models whose favoured status was, after all, to significant extent because they fitted the data that has now shifted, one can construct new models or otherwise modify existing models or theory of gravity to match the current constraints, e.g.~Refs.~\cite{Kallosh:2025rni,Gialamas:2025kef,Brahma:2025dio,Dioguardi:2025vci,Salvio:2025izr,Berera:2025vsu,Dioguardi:2025mpp,Rehman:2025fja,Gao:2025onc,He:2025bli,Yin:2025rrs,Gialamas:2025ofz,Yogesh:2025wak}. 

Here we argue that the shift has substantially benefited curvaton models \cite{Mollerach:1989hu,Linde:1996gt,Enqvist:2001zp,Lyth:2001nq,Moroi:2001ct}, including their simplest incarnation using two quadratic potentials \cite{Bartolo:2002vf,Byrnes:2014xua}. We show that in the usual massless curvaton limit, this model gives the same locus of predictions in the $n_{\rm s}$--$r$ plane as the power-law single-field potentials $\phi^p$. These are often used as a benchmark in assessing observational constraints, but with an upper limit now well below $p=1$, this form can be challenging to motivate from fundamental physics considerations (the best-known option is the monodromy models \cite{Silverstein:2008sg,McAllister:2014mpa}, whose most popular value $p=2/3$ is also now borderline excluded). There is also the question of how to address the non-analyticity at $\phi=0$. We will argue that the curvaton model provides a compelling alternative explanation of current constraints. This is in sharp contrast to the position after the {\it Planck} 2018 results \cite{Planck:2018vyg}, where it appeared the curvaton-dominated limit of these models was comfortably excluded (see their Fig.~28) and no model parameters gave a good fit.

\section{The simplest curvaton model}

\subsection{Model parameters and density perturbations}

We restrict our analysis to the simplest curvaton model \cite{Bartolo:2002vf}. This features two massive non-interacting fields with potential
\begin{equation}
V(\phi,\sigma) = \frac{1}{2}m_\phi^2 \phi^2+ \frac{1}{2}m_\sigma^2 \sigma^2 \,,
\end{equation}
and a flat field-space metric. The inflaton $\phi$ is responsible for driving inflation via its potential energy, while the curvaton $\sigma$ acquires perturbations during inflation that subsequently convert to adiabatic when the curvaton decays to normal matter after inflation ends. We will closely follow the detailed analysis we made in Ref.~\cite{Byrnes:2014xua}, which allows an arbitrary admixture of inflaton and curvaton contributions to the observed adiabatic power spectrum. A similar recent analysis looking at other choices of the inflaton potential was made in Ref.~\cite{Choi:2024ruu}. 

The number of $e$-foldings of inflation from field values $\phi$ and $\sigma$ to the end of inflation is given by 
\begin{equation}
N  = 2\pi \frac{\phi^2 + \sigma^2}{m_{\rm Pl}^2},
\end{equation}
where $m_{\rm Pl}$ is the (non-reduced) Planck mass. By the time observable scales are leaving the horizon, the curvaton field takes a small and effectively constant value $\sigma_*$, which is one of the model parameters. Since this value is small compared to $m_{\rm Pl}$ (the inflating curvaton having been observationally excluded \cite{Byrnes:2014xua}) its impact on the above equation is subdominant to the already tiny contribution from the inflaton field value at the end of inflation, which we have also omitted. Sometimes $\sigma_*$ is estimated using the stochastic prediction for a spectator field in de--Sitter space, which gives $\langle \sigma_*^2\rangle\sim H_*^4/m_\sigma^2$, but the Hubble parameter varies too rapidly during quadratic inflation for this approach to work \cite{Hardwick:2017fjo}.

We consider the complete range from negligible to full curvaton contribution to the total power spectrum, given by
\begin{equation}\label{ps}
{\cal P}^{\rm total}_{\zeta}= {\cal P}^{\phi}_{\zeta}+{\cal P}^{\sigma}_{\zeta}\,.
\end{equation}
We can parametrize the inflaton contribution to the total power spectrum as
\begin{equation}\label{pzeta}
{\cal P}^{\phi}_{\zeta} = \frac{m_{\phi}^2}{m_{\rm single}^2} {\cal P}^{\rm total}_{\zeta}.
\end{equation}
Here $m_{\rm single}$ is the mass that the inflaton would need if it were to give the correct amplitude of perturbations in the single-field case; in a scenario where both fields contribute this is an upper limit to the actual inflaton mass $m_\phi$.  It is determined by
\begin{eqnarray}
{\cal P}_{\rm single} &=& \frac{8 V_{\rm single}}{3 m_{\rm Pl}^4 \epsilon_{\rm single}}\bigg|_{*}\\
&=& \frac{4 m_{\rm single}^2 \phi_{\rm single}^2}{3 m_{\rm Pl}^4} \,2N\bigg|_{*} \,,
\end{eqnarray}
where * refers to the parameter value when observable scales crossed the Hubble radius during inflation, $V_{\rm single}= m_{\rm single}^2 \phi_{\rm single}^2/2$,  and 
\begin{equation}
\epsilon_{\rm single} \equiv \frac{m_{\rm Pl}^2}{16 \pi} \,  \left( \frac{V'}{V}\right)^2=\frac{1}{2N_*}
\end{equation} 
in the single-field model. Taking the observed amplitude as \cite{AtacamaCosmologyTelescope:2025blo} 
\begin{equation} \label{ampNorm}
{\cal P}_\zeta^{\rm obs} \sim 2.1\times 10^{-9}\,,
\end{equation}
we obtain 
\begin{equation}
\frac{m_{\rm single}^2}{m_{\rm Pl}^2}=4.9\times10^{-9}\frac{1}{N_*^2}.
\end{equation}
The ratio $f \equiv m_{\phi}^2/m_{\rm single}^2$ (with $0 < f \leq 1$)\footnote{Not to be confused with our previous use of the symbol $f$ as the decay fraction of inflatons to curvatons in Ref.~\cite{Byrnes:2016xlk}.} will appear throughout in our expressions as a measure of the fractional contribution of the inflaton to the total power spectrum in each model. 

The curvaton contribution to the power spectrum depends on the ratio of curvaton to background energy density at the time the curvaton decays into the thermal bath:
\begin{equation}\label{rdec}
r_{\rm dec} \equiv \frac{3 \rho_\sigma}{4 \rho_\gamma+3 \rho_\sigma}\bigg|_{\rm decay} \,,
\end{equation}
where we assumed that the inflaton has fully decayed into radiation before the curvaton decays. Equation~(\ref{rdec}) is defined so as to provide a unified expression for the curvaton perturbation in the regimes of radiation and curvaton domination at the time of decay, which is \cite{Lyth:2002my}
\begin{equation}\label{Psigma}
{\cal P}_\zeta^{\sigma} = \frac{r_{\rm dec}^2}{9 \pi^2}\frac{H_*^2}{\sigma_*^2}\,.
\end{equation}
We use the normalization amplitude Eq.~(\ref{ampNorm}) to fix the ratio $r_{\rm dec}^2 H_*^2/\sigma_*^2$ and obtain  
\begin{equation}\label{rdec-our}
r_{\rm dec}^2  = 2.8 \times 10^{-7}  \left(1-f\right)  \frac{\sigma_{*}^2}{ m_{\phi}^2 N_*} \,,
\end{equation}
where henceforth $N_*$ is the $e$-foldings number at which the {\it Planck} normalization scale $0.05\, {\rm Mpc}^{-1}$ crosses the Hubble radius during inflation. The required $r_{\rm dec}$ value fixes the decay timescale of the curvaton. Evaluating Eq.~(\ref{rdec}) requires knowledge of the full curvaton evolution, but in practice we will only use $r_{\rm dec}$ via Eq.~(\ref{rdec-our}) as a constraint on model parameters by requiring that it takes the physically realisable values $0<r_{\rm dec}<1$. This also allows a simple generalisation to models where the inflaton decays partly to curvatons which later themselves decay into the thermal bath \cite{Byrnes:2016xlk}. We will see that the lower bound is tightened by the constraint on local $\fnl$.

\subsection{Parametrization of the number of $e$-foldings}\label{nefold}

To impose accurate constraints we need to identify the correct number of $e$-foldings corresponding to the $k_*=0.05 \, {\rm  Mpc}^{-1}$ pivot scale at which observables are evaluated. With $V_{\rm hor}$ the potential when the pivot scale crossed the horizon during inflation, $\rho_{\rm end}$ the energy density at the end of inflation, and $\rho_{\rm reh}$ the energy density at the end of reheating, this is given by \cite{Liddle:2003as,Byrnes:2014xua}
\begin{equation}\label{Nhor}
N_{\rm *}=58+\frac{1}{4}\ln (rf)+ \frac{1}{4}\ln\frac{V_{\rm hor}}{\rho_{\rm end}}+\frac{1}{12}\ln\frac{\rho_{\rm reh}}{\rho_{\rm end}}\,,
\end{equation}
where all quantities are as in single-field models. The factor $f$ is needed because it modifies the normalization of $V_{\rm hor}$ relative to the observed power spectrum amplitude when the curvaton perturbations are non-negligible\footnote{This factor was omitted in Ref.~\cite{Byrnes:2014xua}, but the error did not propagate.}, but where we need to include the epoch of curvaton reheating in addition to the usual inflaton reheating. We parametrize the total amount of reheating $e$-foldings, given by the last term of Eq.~(\ref{Nhor}), as $N_{\rm matter}$ which includes the reheating of both the inflaton and the curvaton, obtaining 
\begin{equation}
\frac{1}{12}\ln\frac{\rho_{\rm reh}}{\rho_{\rm end}} =- \frac{1}{4}N_{\rm matter}\,.
\end{equation}
For the quadratic inflaton, the middle two terms in Eq.~(\ref{Nhor}) combine into a term which measures the inflaton mass relative to the single-field limit (i.e.\ those terms cancel in the single-field case), giving
\begin{equation}\label{Nstar}
N_* =  58+\frac{1}{4} \ln f-\frac{1}{4} N_{\rm matter}.
\end{equation}
We note that this relation sets a firm upper limit of $58$ as the number of $e$-foldings corresponding to the {\it Planck} pivot scale (see also Ref.~\cite{German:2022sjd}). 

The appropriate choice of $N_*$ remains a significant uncertainty in inflation modelling \cite{Liddle:2003as}. $N_{\rm matter}$ has quite a wide plausible range, for instance $N_{\rm matter} = 0$ implies instant reheating and no curvaton domination, while positive values allow for both the period of reheating after inflation and any subsequent period of curvaton domination.  $N_{\rm matter} > 16$ would ensure reheating at less than $10^{11} {\rm GeV}$ to avoid overproduction of gravitinos, while $N_{\rm matter} \lesssim 40$ is necessary to ensure reheating before electro-weak symmetry breaking. Reducing the energy scale of inflation is more efficient at reducing $N_*$ and permitting the curvaton to complete reheating before the electroweak scale is reached, whilst allowing circa 15 e-foldings of expansion for the curvaton to dominate after starting with a plausible energy density a million times smaller than the inflaton's leads to $N_*\simeq38$, showing that even $N_* \lesssim 40$ can be consistent with models of electroweak baryogenesis. Overall, values of $N_*$ in the range $45$ to $55$ are reasonable, with the lower end corresponding to the usual curvaton-dominated regime and negligible inflaton fluctuations.

\begin{figure*}[t]
\centering 
\includegraphics[width=.75\textwidth]{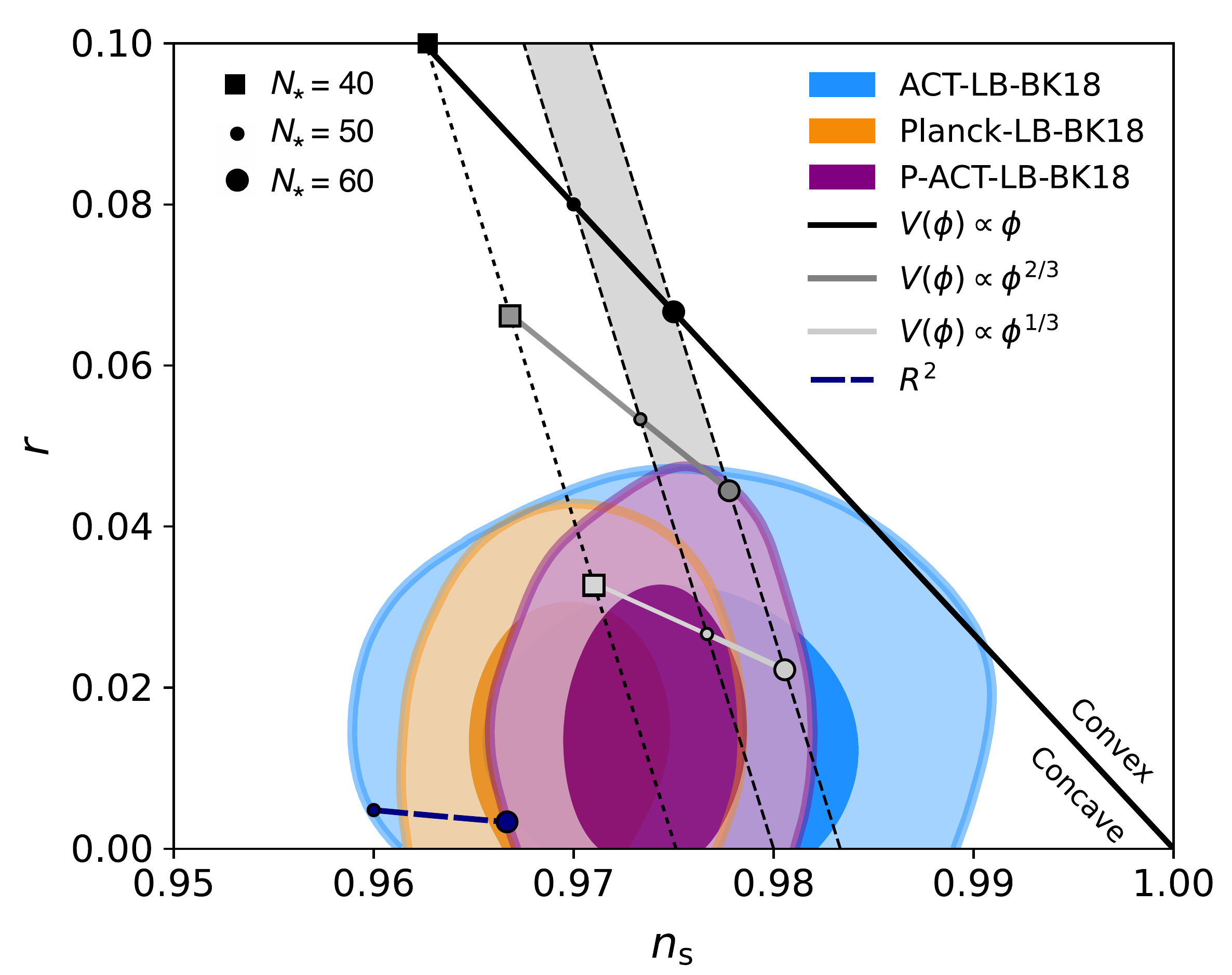}
\caption{\label{fig:ACTfig}  Observational constraints in the $n_{\rm s}$--$r$ plane from Ref.~\cite{AtacamaCosmologyTelescope:2025nti}. The combined {\it Planck}, ACT and BICEP/Keck CMB spectral data, enhanced by CMB lensing and by baryon acoustic oscillation data from DESI-DR1, are shown in purple (see also Figure 59 in the final version of their article for an update to DESI-DR2). The three dashed lines show the power-law potentials for $e$-folding choices $N_* = 40$, $50$, and $60$, which are also the predictions of our curvaton model in the massless curvaton limit. The sets of symbols, from top to bottom, correspond to $f = 1/2$, $1/3$, and $1/6$. [Adapted from Fig.~10 of Ref.~\cite{AtacamaCosmologyTelescope:2025nti}, under Creative Commons BY 4.0 License.]}
\end{figure*}

\subsection{Observables: Linear power spectrum}

We can now make predictions for model observables: the spectral index $n_{\rm S}$, tensor-to-scalar ratio $r$, and non-Gaussianity parameter $\fnl$. The slow-roll parameters are defined by 
\begin{eqnarray}
\epsilon=-\frac{\dot{H}}{H^2}\simeq \frac12 \left(\frac{dV/d\phi}{3H^2}\right)^2 ; \quad \\ \eta_\phi = \frac{d^2 V/d\phi^2}{3H^2}\,;\;\;\eta_\sigma = \frac{d^2V/d\sigma^2}{3 H^2} \,. \nonumber
\end{eqnarray}
If there is a single period of inflation, $\epsilon \simeq \eta_\phi \simeq 1/2N_*$.\footnote{This corrects a typo in Ref.~\cite{Byrnes:2014xua}, which fortunately was not propagated to any of the results.}

The predicted deviation from scale invariance is \cite{Wands:2002bn},
\begin{equation}\label{ns}
n_{\rm S}-1= \left(1-f\right) (-2\epsilon+2\eta_{\sigma}) +f\left(-6\epsilon+2 \eta_{\phi}\right).
\end{equation}
This expression allows for ready distinction between the contribution of the inflaton versus curvaton for each model spectrum, as a function of the inflaton contribution $f$. It reduces to each of these two regimes for $f \sim 1$ and $f \ll 1$ respectively. We further note that $n_{\rm S}$ has no dependence on $\sigma_*$ at leading order in slow-roll. Lastly, Eq.~(\ref{ns}) has an implicit dependence on the number of $e$-foldings and the chosen pivot scale, which we presented in Section~\ref{nefold}.

The tensor-to-scalar ratio $r$ is given by \cite{Byrnes:2014xua}
\begin{equation}\label{r_tens}
r =16\, \epsilon f \,,\\
\end{equation}
also parametrized by the relative contribution of the inflaton mass, and without further dependence on curvaton parameters. As $m_\phi$ is varied from zero to $m_{\rm single}$, the prediction in the $n_{\rm S}$--$r$ plane interpolates linearly between the curvaton-dominated and inflaton-dominated regimes.

The standard reference guide in discussing observational constraints on inflation is the power-law potential $V(\phi) \propto \phi^p$, whose predictions in the $n_{\rm s}$--$r$ plane are seen in Fig.~\ref{fig:ACTfig} for several $N_*$ values. While early inflation modelling envisaged an even integer power such as $p=2$ or $4$, observational constraints, particularly the non-detection of $r$, have now driven the limit on $p$ well below one, making the justification and interpretation of the potential much harder. Even the most-preferred monodromy value of $p=2/3$ \cite{Silverstein:2008sg,McAllister:2014mpa} is under strong pressure, especially for lower $N_*$ values. But remarkably, in the limit of negligible curvaton mass, the curvaton model predicts exactly the same locus in the $n_{\rm s}$--$r$ plane, as we now show.

Taking $\eta_\sigma$ to be negligible, the curvaton model predictions of Eqs.~(\ref{ns}) and (\ref{r_tens}) simplify to
\begin{equation}
n_{\rm s} - 1 = -(1+f)\frac{1}{N_*} \quad ; \quad r = \frac{8f}{N_*} \,.
\end{equation}
For comparison, a single-field model with $V \propto \phi^p$ gives \cite{Liddle:1992wi}
\begin{equation}
n_{\rm s} - 1 = - \left(1+\frac{p}{2} \right) \frac{1}{N_*} \quad ; \quad r = \frac{4p}{N_*} \,, 
\end{equation}
which yields the dashed lines shown in Fig.~\ref{fig:ACTfig}. Hence, we have the simple correspondence
\begin{equation}
    f \longleftrightarrow \frac{p}{2} \,.
\end{equation}
This lets the curvaton model match any $p$ in the range $0$ to $2$, passing between curvaton and inflaton domination of the generated perturbations respectively. In particular, the curvaton model can exactly mimic the combined $(n_{\rm s},r)$ values of $p<1$ while using only simple quadratic potentials. 

The current observational constraints on $r$ imply \mbox{$f\lesssim1/3$} (at 95\% confidence), becoming somewhat stronger if $N_*$ is well below 60. Hence, the curvaton perturbations must dominate, but there is still space for a significant (up to 33\%) contribution from the inflaton field to the total power spectrum. The curvaton-dominated regime $f \rightarrow 0$, corresponding to $p \rightarrow 0$, gives a good account of the data for the most plausible values of $N_*$.

Despite the match in the $n_{\rm s}$--$r$ predictions, the running of the spectral index will not be identical, as a change in $N_*$ changes the value of $m_{\rm  single}$ and hence $f$.\footnote{We thank David Wands for pointing this out to us.} The runnings $\alpha \equiv dn_{\rm s}/d\ln k$ for the $\phi^p$ model \cite{Kosowsky:1995aa,Chiba:2014gfa} and the curvaton model \cite{Byrnes:2006fr} are given, respectively, by
\begin{eqnarray}
    \alpha_{\phi^p}&=&-(n_s-1)^2-\frac{r}{8}(n_s-1) \,, \\
    \alpha_{\rm curv}&=&-(n_s-1)^2-\frac{r}{4}(n_s-1)-\frac{r^2}{32} \,.
\end{eqnarray}
Their difference
\begin{equation}
    \alpha_{\phi^p}-\alpha_{\rm curv}=\frac{f(1-f)}{N_*^2} \,,
\end{equation}
vanishes in the pure inflaton or pure curvaton limits as expected, and given current limits the difference cannot be larger than $\sim 10^{-4}$, an order of magnitude below the expected sensitivity of near-future experiments \cite{BigBoss:2011xpw,Kohri:2013mxa,Munoz:2016owz}.

If the curvaton mass is non-negligible, $n_{\rm s}$ is shifted to larger values (greater than unity is even possible); see e.g.~Fig.~6 of Ref.~\cite{Byrnes:2014xua}. This worsens the data fit, but the observations only limit $m_\sigma$ to be less than of order $m_\phi$ which is not very constraining. For example, in the curvaton-dominated limit ($f=0$), the spectral index is the same if $N_*=50$ with $\eta_\sigma \ll \eta_\phi$ or if $N_*=40$ with $\eta_\sigma=\eta_\phi/5$, which corresponds to $m_\sigma\simeq0.45m_\phi$.

\subsection{Observables: Non-Gaussianity}\label{sec:NonG}

The curvaton scenario generates non-Gaussianity with the local shape, parametrized by the usual $\fnl$ parameter whose value is \cite{Lyth:2002my,Sasaki:2006kq,Enqvist:2013paa}.
\begin{equation}\label{fnl}
f_{\rm NL} = \frac{5}{12} \left(1-f\right)^2 \left(\frac{3}{r_{\rm dec}}-4-2 r_{\rm dec}\right).
\end{equation}
Note that this expression is independent of the curvaton mass, which does not appear in the expression for $r_{\rm dec}$. The non-Gaussianity predictions therefore also depend on only two model parameters, but in a plane orthogonal to the two that determine $n_{\rm S}$.

In the limit of $m_\phi\ll m_{\rm single}$ this reduces to the standard curvaton result, while the value is suppressed if the (nearly) Gaussian inflaton perturbations also contribute to the total power spectrum. The limit on $f$ above indicates a maximum suppression of about a factor of $(2/3)^2 \simeq 0.4$.

Taking the {\it Planck} 95\% confidence observational upper limit of $\fnl<9.3$ \cite{Planck:2019kim} leads to $r_{\rm dec}>0.11$ in the curvaton limit and $r_{\rm dec}>0.05$ if the inflaton contributes 33\% of the perturbations. We note that values of $r_{\rm dec}$ quite close to but not equal to one require rather fine-tuned parameters, and hence $r_{\rm dec}\simeq1$ should arguably be considered the natural expectation \cite{Hardwick:2015tma,Torrado:2017qtr}. We therefore expect $\fnl=-5/4$ if the curvaton dominates, a value which SPHEREx, which launched in March 2025, may be able to distinguish from Gaussianity in the next few years \cite{SPHEREx:2014bgr,Heinrich:2023qaa}. A detection of $\fnl<-5/4$ would completely rule out the quadratic curvaton scenario independently of the inflaton potential, whilst a constraint of e.g.~$|\fnl|<0.5$ would strongly disfavour the curvaton scenario over single-field scenarios which gives rise to $\fnl\simeq0$. A detection of $\fnl>0$ would rule out single-field inflation and put tight constraints on $r_{\rm dec}$. For example if SPHEREx detected $2<\fnl<4$ and if constraints on $r$ also significantly strengthened, then this would require $0.21<r_{\rm dec}<0.32$ for the quadratic curvaton scenario to remain viable.

\section{Conclusions}

The significant shift in $n_{\rm s}$ prompted by the ACT DR6 results \cite{AtacamaCosmologyTelescope:2025blo} has restored the viability of simple curvaton models. We have uncovered an exact correspondence between the observational predictions of the simplest curvaton model \cite{Bartolo:2002vf} and the single-field $\phi^p$ potential that is widely used as a benchmark, with the power-law $p$ replaced by the admixture $f$ of inflaton and curvaton contributions to the perturbations. Moreover, the curvaton model uses only simple quadratic potentials, yet can give predictions that mimic $\phi^p$ models with $p \ll 1$ that can be otherwise hard to motivate. 

The current limit $p \lesssim 2/3$ corresponds to $f \lesssim 1/3$, excluding the inflaton-dominated regime while still permitting a substantial admixture. The curvaton-dominated regime $f \rightarrow 0$ is also a good fit to the data, especially if the $e$-folding number of the perturbation pivot scale is towards the lower end of our fiducial range $N_* = 45$ to $55$. Addition of the newest South Pole Telescope SPT-3G D1 dataset maintains this conclusion in favor of the curvaton model. The data compilation CMB-SPA\,+\,DESI (Table VI of Ref.~\cite{SPT-3G:2025bzu}) yields $n_{\rm s} = 0.9728 \pm 0.0027$, though with some caveats on combining these datasets into a joint analysis.

While we have focused on the simplest implementation of the curvaton scenario \cite{Bartolo:2002vf,Byrnes:2014xua}, clearly other variants can also be expected to give a good fit to current data. Non-Gaussianity provides a challenging but feasible future route to distinguishing these paradigms from the single-field models.

\section*{Acknowledgements}
We thank Renata Kallosh and Andrei Linde for comments. Also, we thank Charlotte Wood for pointing out an error in the original version of Eq.~(\ref{Nhor}), and David Wands for important insights concerning the spectral index running. CB is supported by STFC grants ST/X001040/1 and ST/X000796/1. M.C and A.R.L. were supported by the Funda\c{c}\~{a}o para a Ci\^encia e a Tecnologia (FCT) through the research grants UIDB/04434/2020 and UIDP/04434/2020. M.C.\ acknowledges support from the FCT through the Investigador FCT Contract No.\ CEECIND/02581/2018 and POPH/FSE (EC), and A.R.L.\ through the Investigador FCT Contract No.\ CEECIND/02854/2017 and POPH/FSE (EC). This article/publication is based upon work from COST Action CA21136 -- ``Addressing observational tensions in cosmology with systematics and fundamental physics (CosmoVerse)'', supported by COST (European Cooperation in Science and Technology). 

\bibliographystyle{apsrev4-2}  
\bibliography{references}                                        
\end{document}